\documentstyle[multicol,aps,prl,psfig]{revtex} 

\renewcommand{\narrowtext}{\begin{multicols}{2} \global\columnwidth20.5pc}
\renewcommand{\widetext}{\end{multicols} \global\columnwidth42.5pc}
\def\al{\alpha}
\def\be{\beta}

\def\de{\delta}
\def\ep{\epsilon}

\def\et{\eta}

\def\ka{\kappa}
\def\la{\lambda}

\def\De{\Delta}

\def\La{\Lambda}

\def\cl{{\cal L}}
\def\fr#1#2{{{#1} \over {#2}}}
\def\frac#1#2{{\textstyle{{#1}\over {#2}}}}
\def\half{{\textstyle{1\over 2}}}
\def\prt{\partial}

\def\lsim{\mathrel{\rlap{\lower4pt\hbox{\hskip1pt$\sim$}}
    \raise1pt\hbox{$<$}}}
\def\gsim{\mathrel{\rlap{\lower4pt\hbox{\hskip1pt$\sim$}}
    \raise1pt\hbox{$>$}}}
\def\sqr#1#2{{\vcenter{\vbox{\hrule height.#2pt
         \hbox{\vrule width.#2pt height#1pt \kern#1pt
         \vrule width.#2pt}
         \hrule height.#2pt}}}}

\def\aaa{a} 
\def\ppp{T} 
\def\kkk{\ka}

\newcommand{\beq}{\begin{equation}}
\newcommand{\eeq}{\end{equation}}
\newcommand{\bea}{\begin{eqnarray}}
\newcommand{\eea}{\end{eqnarray}}
\newcommand{\rf}[1]{(\ref{#1})}
 
\begin{document}

\title{Off-shell structure of the string sigma model}
\author{V.\ Alan Kosteleck\'y,$^a$ Malcolm J.\ Perry,$^b$ 
and Robertus Potting$^{a,c}$} 
\address{$^a$Physics Department, Indiana University, Bloomington, 
Indiana 47405, U.S.A.} 
\address{$^b$D.A.M.T.P., Silver Street, Cambridge University, 
Cambridge, England} 
\address{$^c$Universidade do Algarve, U.C.E.H., Campus de Gambelas, 
8000 Faro, Portugal} 
\date{preprint IUHET 412 (1999)}
\maketitle

\begin{abstract}
The off-shell structure of the string sigma model is investigated. 
In the open bosonic string,
nonperturbative effects are shown to depend crucially 
on the regularization scheme.
A scheme retaining the notion of string width 
reproduces the structure of Witten's string field theory
and the associated unconventional gauge transformations.
\end{abstract}

\narrowtext

Numerous approaches to string theory exist.
Many involve the on-shell structure of strings,
thereby providing information about topics such as scattering amplitudes.
However,
the off-shell structure of string theory is also of interest.
In the case of particle physics,
the transition between on-shell and off-shell physics
is implemented by the step from relativistic quantum mechanics
to particle quantum field theory,
and this ultimately provides insights 
about topics such as the structure of the vacuum.
Understanding off-shell string physics could similarly 
lead to substantial physical insights.

Direct construction of a field theory for strings is possible
in certain cases
\cite{ew}
and provides some evidence for nontrivial vacuum structure
\cite{kskp},
but this program has proven difficult to implement for general strings.
A more practical approach to off-shell string structure 
is to obtain the equations of motion for the particle fields 
associated with the string modes
and subsequently to reconstruct the corresponding action.
In the string sigma model,
for example,
background particle fields can be introduced 
and their equations of motion extracted 
by imposing world-sheet conformal invariance 
\cite{r1}.
If the fields $g^k$ are viewed as couplings on the world sheet,
this procedure amounts to imposing the vanishing of 
the renormalization-group beta functionals $\be^j$.
The $\be^j$ are expected to originate from variation 
of a background-field action $I$,
\beq
\be^j=G^{jk} \de I/\de g^k,
\label{beI}
\eeq
for a suitable metric $G^{jk}$
\cite{pz}.
This suggests that $I$ could be an appropriate tree-level action 
in a field-theory formulation for strings.

Although the literature on the string renormalization group
and the beta functionals is extensive 
(see, for example, Refs.\
\cite{ds,bm,klsu,hlp,ft,acny,ao,yw,bs,bny,es,bkp,kb}),
results for the action $I$ 
have suffered from a degree of arbitrariness
\cite{at}
because the calculations are typically performed in a fixed gauge 
and subsequently converted to covariant form
using the equations of motion to redefine the fields.
This process introduces free parameters
and leads to nonlocal or even singular actions.
It also has the disadvantage that information about
the static potential in $I$ and hence 
about nontrivial string vacua is lost, 
as nonderivative terms can be arbitrarily replaced by derivatives.

In this work, 
we investigate a background-field action without fixing the gauge.
In a general weak-field expansion,
the beta functionals $\be^j$ of the fields $g^j$ at scale $t$
can be written 
$\be^j\equiv dg^j(t)/dt=\la_jg^j+\al_{kl}^jg^kg^l + \ldots$
(no sum on $j$).
This equation has the solution
\cite{klsu}
\beq
g^j(t)=e^{\la_jt}g^j(0)+(e^{(\la_k+\la_l)t}-e^{\la_jt})
{\al_{kl}^jg^k(0)g^l(0)\over\la_k+\la_l-\la_j}.
\label{git}
\eeq
The idea here is to establish explicitly
the behavior of renormalized fields 
as a function of the cutoff parameter $\aaa$ 
in a two-dimensional field theory,
and then by comparison with Eq.\ \rf{git}
to deduce the corresponding beta functionals
and hence the field-theory action.

For definiteness,
we consider open bosonic strings.
It suffices for our purposes to investigate couplings involving 
a background of tachyons $\ppp$ and massless vector fields $A^\mu$,
so without loss of generality we can take as the starting point
the standard action 
\bea
S&=&{1\over 4\pi\al'}
\int_{y>0}dx\,dy\,\prt_a X_\mu
\prt^a X^\mu 
+{g\over 2\sqrt2}\int dx {\cl}_B ~,
\nonumber\\
{\cl}_B&=& 
\int d^Dk\,e^{ik\cdot X} 
\left({\sqrt{2\al'}\over\aaa} \ppp(k)+i\fr{dX_\mu}{dx}A^\mu(k)\right),
\label{standard}
\eea
where $\aaa$ can be regarded as the cutoff  
in the two-dimensional field theory.
Given a classical background $X_0^\mu(x,y)$
satisfying the equations of motion,
we write $X^\mu$ as $X^\mu=X_0^\mu+\xi^\mu$
and perform a weak-field expansion in the effective action.
The boundary propagator relevant for calculation of expectation values is
$\langle \xi^\mu(x_1)\xi^\nu(x_2) \rangle
=-2\al'\et^{\mu\nu}\ln |\De x|$,
where $\De x \equiv x_1 - x_2$.

We perform a calculation of the linear and quadratic terms
in the effective action using two distinct types of regulator.
One involves direct regularization of the correlator.
The second involves a regularization of the effective action
that cannot be derived from a regularized correlator.
Both calculations are nonperturbative in $\al'$.
In what follows,
we first present the calculations and then discuss the results.

To begin,
consider a one-parameter family of correlator regularizations
defined by
\beq
\langle \xi^\mu(x_1)\xi^\nu(x_2) \rangle_\kkk
=-{2\over\kkk}\al'\et^{\mu\nu}\ln (|\De x|^\kkk+\aaa^\kkk)
\quad ,
\label{regkkk}
\eeq
where $\kkk$ is a positive parameter.
In the $\kkk\rightarrow\infty$ limit,
this corresponds to a step-function cutoff:
\beq
\langle \xi^\mu(x_1)\xi^\nu(x_2) \rangle_\infty
=-2\al'\et^{\mu\nu}\left\{\begin{array}{ll}
\ln |\De x|,&|\De x|>\aaa;\\
\ln\aaa,&|\De x|<\aaa.
\end{array} \right.
\label{regkkkinfty}
\eeq
The calculation is presented explicitly here only in this limit.
For simplicity in what follows, we set $\al' = 1$.

Expanding the effective action to second order in 
the string coupling $g$ suffices to extract
the linear and quadratic terms.
This corresponds to coupling up to two background fields 
to the string world sheet.
The effective action involves divergent integrals
over the coupling points $x_1$ and $x_2$.
Following the standard procedure,
we Taylor-expand $x_2$-dependent factors around $x_1$,
regularize using Eq.\ \rf{regkkkinfty},
and perform the $x_2$ integral.
Comparing coefficients of $dX_\mu/dx$
then allows the extraction of
the renormalized fields as a function of $\aaa$:
\bea
\ppp_{\infty,\aaa}(k)&=&\aaa^{k^2-1}\bigl[\ppp(k)
+ 2\int_k
\left( 
\fr{k_1\cdot k_2 \ppp(k_1)\ppp(k_2)}{2k_1\cdot k_2+1}
\right.
\nonumber \\
&&
\left.
+\fr{(k_1^\nu k_2^\mu-k_1\cdot k_2\et^{\mu\nu})A_\mu(k_1)A_\nu(k_2)}
{2k_1\cdot k_2-1}
\right)\bigr] ,
\nonumber \\
A^\mu_{\infty,\aaa}(k)&=&
\aaa^{k^2}\bigl[ A^\mu(k)
\nonumber \\
&&
- 4\int_k
\fr{( k_1^\nu k_2^\mu - k_1\cdot k_2 \et^{\mu\nu}) \ppp(k_1)A_\nu(k_2)}
{2k_1\cdot k_2+1}
\bigr] ,
\label{Aren}
\eea
where
$\int_k \equiv {g\over2} \int dk_1\,dk_2\,\de^D(k_1+k_2-k)$.
The nonperturbative nature of the calculation
requires fields at all mass levels 
to be included for consistency.
However,
for simplicity throughout this work 
we truncate the full expressions to 
terms containing only $\ppp$ and $A^\mu$.

As expected,
the form of Eq.\ \rf{Aren} matches 
that of the general solution \rf{git}.
With $a\equiv e^{-t}$ we reproduce the factors $e^{\la_jt}$, 
but the terms $e^{(\la_k+\la_l)t}$ are absent.
This is because the infrared cutoff has been removed
in the calculation,
while the only physically relevant scale 
would be the ratio between ultraviolet and infrared cutoffs
\cite{klsu}.
Explicitly reintroducing an infrared cutoff reinstates the missing terms.
We can thus identify
$\la^{\ppp(k)}=1-k^2$
and 
$\la^{A^\mu}=-k^2$.
This permits the identification of the coefficients $\al^j_{kl}$ 
in Eq.\ (\ref{git}).

A total derivative
$k^\mu\La(k)$ with arbitrary $\La$ can be added
to the second of Eqs.\ (\ref{Aren}) 
because it vanishes in the integral over the boundary.
This freedom provides some flexibility
in the choice of $\la^j$ and $\al^j_{kl}$.
Decomposing $A^\mu = A^\mu_\perp + A^\mu_\parallel$,
with $A^\mu_\parallel \equiv k^\mu (k\cdot A)/k^2$,
reveals that the values of  
$\la^{A^\mu_\parallel}$ and $\al_{kl}^{A^\mu_\parallel}$
remain unprescribed.
An attractive choice at the linear level
is $\La(k)=-k\cdot A(k)$.
As shown below,
this decouples the longitudinal part of $A_\mu(k)$
in the beta functionals,
makes $\la^{A^\mu_\parallel}$ vanish,
and leads to
a gauge-invariant equation of motion for the stationary points
of the renormalization-group flow.
At the nonlinear level, 
there is also always a unique way 
(varying with the regularization scheme) 
to maintain gauge invariance by correctly choosing $\La(k)$.

Careful comparison of Eqs.\ (\ref{Aren}) with Eq.\ (\ref{git}), 
keeping $A^\mu_{\perp}$ and $A^\mu_{\parallel}$ separate 
and replacing the discrete sums over $j$ and $k$ in Eq.\ (\ref{git})
by integrals over $k_1$ and $k_2$,
permits the extraction of the desired beta functionals
\bea
\be_{\infty,\ppp(k)}&=&-(k^2-1)\ppp(k)
-2\int_k
\bigl[
k_1\cdot k_2
\ppp(k_1)\ppp(k_2)
\nonumber\\ &&
+(k_1^\nu k_2^\mu-k_1\cdot k_2\et^{\mu\nu})
A_{\perp\mu}(k_1)A_{\perp\nu}(k_2)
\bigr] ,
\nonumber\\
\be_{\infty,A^\mu_\perp(k)}&=&
-k^2A^\mu_\perp(k)
\nonumber\\ &&
+4\int_k
(k_1^\nu k_2^\mu-k_1\cdot k_2\et^{\mu\nu})
\ppp(k_1)A_{\perp\nu}(k_2) ,
\nonumber\\
\be_{\infty,A^\mu_\parallel(k)}&=&{\cal O}(g)
\quad .
\eea
As explained above,
the latter beta functional is 
chosen to vanish at order $g^0$ but is otherwise arbitrary.

These beta functionals cannot be obtained directly from an action.
However, 
it suffices to obtain a relation of the form (\ref{beI})
for some functionals $G^{jk}$.
This can be achieved to order $g$ with
\bea
G^{\ppp(k),\ppp(k_1)}&=&
-\de(k-k_1)-2g\ppp(k-k_1)
\quad ,
\nonumber\\
G^{A_\perp^\mu(k),A_\perp^\nu(k_1)}&=&
\left[-\de(k-k_1) -2g\ppp(k-k_1)\right] \et^{\mu\nu}
\quad ,
\label{Gval}
\eea
and vanishing other $G^{jk}$. 
In covariant form,
the corresponding lagrangian in coordinate space becomes
\bea
{\cl}_\infty &=&
\half\ppp(\prt^2+1)\ppp
- \frac 14 F_{\mu\nu}F^{\mu\nu}
+ \frac 2 3 g\ppp^3 
\nonumber\\
&&
-g \ppp \prt_\mu\ppp \prt^\mu\ppp
+ \half g\ppp F_{\mu\nu}F^{\mu\nu}.
\label{I1}
\eea
This lagrangian is invariant under the usual
gauge transformations $\de\ppp(k)=0$, $\de A^\mu(k)=\prt^\mu\ep$.

Next,
we repeat the above calculation in an alternative 
regularization scheme.
The expectation values for coincident points are taken to be the same
as before,
so the critical exponents $\la^{\ppp(k)}$ and $\la^{A^\mu(k)}$
and the analysis of the linear terms in the beta functionals
remains unchanged.
However,
for noncoincident points $x_1$, $x_2$ in the quadratic terms 
we now stipulate the brick-wall condition $|\De x|>\aaa$.
This regularizes the integrals over $x_1$, $x_2$ 
by excluding intervals with two points closer than $\aaa$.
The correlators are otherwise unchanged
\cite{fn1}.

The renormalized fields as functions of $\aaa$ now become 
\bea
\ppp_\aaa(k)&=&\aaa^{k^2-1}\bigl[\ppp(k)
-\int_k
\left({\ppp(k_1)\ppp(k_2)\over2k_1\cdot k_2+1}
\right.
\nonumber\\
&&
\qquad \qquad
\left.
- {(2k_1^\nu k_2^\mu-\et^{\mu\nu})A_\mu(k_1)A_\nu(k_2)
\over2k_1\cdot k_2-1}\right)\bigr]
\quad ,
\nonumber\\
A^\mu_\aaa(k)&=&
\aaa^{k^2}\bigl[A^\mu(k)
-2\int_k
\fr{(2k_1^\nu k_2^\mu + \et^{\mu\nu}) \ppp(k_1)A_\nu(k_2)}
{2k_1\cdot k_2+1}
\bigr] .
\nonumber\\
\label{Aren2}
\eea
As before, 
we are free to add $k^\mu\La(k)$
to the second equation,
resulting in an arbitrary beta functional
$\be(A_\parallel^\mu(k))$.
The other beta functionals are
\bea 
\be_{\ppp(k)}&=&-(k^2-1)\ppp(k)
+\int_k
\bigl[\ppp(k_1)\ppp(k_2)
\nonumber\\ &&
\qquad
-(2k_1^\nu k_2^\mu-\et^{\mu\nu})
A_{\perp\mu}(k_1)A_{\perp\nu}(k_2)
\nonumber\\ &&
\qquad
-\et^{\mu\nu}(k^2-1)A_{\parallel\mu}(k_1)A_{\parallel\nu}(k_2)
\nonumber\\ &&
\qquad
-2\et^{\mu\nu}(k^2-1-k_1^2)
A_{\perp\mu}(k_1)A_{\parallel\nu}(k_2)
\bigr] ,
\nonumber\\ 
\be_{A^\mu_\perp(k)}&=&
-k^2A^\mu_\perp(k)
+2\int_k
\bigl[
(k^2-k_1^2+1)\ppp(k_1)A_\parallel^\mu(k_2)
\nonumber\\ &&
\qquad \qquad
+(2k_1^\nu k_2^\mu + \et^{\mu\nu})
\ppp(k_1)A_{\perp\nu}(k_2)\bigr] .
\eea
They obtain directly from the covariant lagrangian
\bea
{\cl}&=&
\half\ppp(\prt^2+1)\ppp
- \frac 14 F_{\mu\nu}F^{\mu\nu}
+ \frac 1 {3!} g\ppp^3 
+ \half g \ppp A_\mu A^\mu
\nonumber\\
&&
+g \ppp \left(
\prt_\nu A^\mu \prt_\mu A^\nu + A^\mu \prt_\mu \prt_\nu A^\nu 
\right)
\quad ,
\label{I2}
\eea
invariant under the modified gauge transformations
\beq
\de\ppp=-g A^\mu \prt_\mu \ep , \quad 
\de A^\mu= (1 + g \ppp )\prt^\mu\ep.
\label{gaugeinv2}
\eeq

The two calculations above confirm that to this order 
the renormalization-group flow is a gradient flow
\cite{fn1a},
albeit involving two different lagrangians (\ref{I1}) and (\ref{I2}).
Both lagrangians are superficially similar,
having cubic couplings and identical kinetic terms. 

It is of particular interest to note that
both lagrangians contain a static $\ppp^3$ coupling. 
This arises in ${\cl}_\infty$
despite the linearity of $\be_{\infty,\ppp(k)}$
in the static $\ppp$ modes,
as a consequence of the existence of the nontrivial $G^{jk}$
required for the existence of the action 
\cite{fn2}.
There are two general types of argument in the literature 
in favor of a purely quadratic tachyon potential
\cite{fn3}.
The first draws conclusions from the beta functional,
which disregards the need for nontrivial $G^{jk}$.
The second derives a trivial scaling behavior for the tachyon field 
using perturbative string theory,
which is insensitive to nonperturbative effects. 

It is also of interest that 
the gauge properties of the two lagrangians \rf{I1} and \rf{I2}
are qualitatively different.
Each term in  ${\cl}_\infty$ 
is invariant under the usual particle gauge transformations.
However,
the gauge invariance of $\cl$ is modified at order $g$
relative to the usual case
\cite{fn4},
and individual terms are not independently gauge invariant.
Indeed,
the static $\ppp A_\mu A^\mu$ coupling 
would be incompatible with the usual particle gauge symmetry.
Note that the issue of the structure of the lagrangian
is potentially of more than academic interest.
For example,
the occurrence of static interaction terms
such as $\ppp A_\mu A^\mu$ 
between tensors and scalars
may trigger spontaneous symmetry breaking
with expectation values not merely for scalars
but also for Lorentz tensors
\cite{kskp}.
If this occurs under favorable circumstances in a realistic string theory,
it might produce an observable spontaneous breakdown 
of Lorentz and CPT invariance 
\cite{ckp}.

The origin of the difference in the two calculations can be traced to
the integration of correlators of the form
$\langle e^{ik_1\cdot\xi(x_1)} e^{ik_2\cdot\xi(x_2)}\ldots\rangle$,
where the ellipsis refers to factors of the form $\prt_{x_j}\xi(x_j)$
indicating coupling to $A^\mu(k_j)$.
The scalar product of this expression with $k_j^\mu$
determines the longitudinal coupling to $A_\parallel^\mu(k_j)$
and arises by differentiation with respect to $x_j$,
i.e., it is a total derivative.
Therefore,
in the first regularization scheme
in which an integration is performed over all real values of $x_j$,
the longitudinal coupling necessarily vanishes
(assuming no contribution from boundary terms).
This argument holds for any scheme based on regularizing the correlator
in which $x_2$ can take arbitrary values relative to $x_1$,
including the scheme of Eq.\ (\ref{regkkk}) for generic $\kkk$.
In contrast,
the brick-wall scheme avoids this requirement
because the regularization excludes 
an interval of width $\aaa$ from the integration.
See Fig.\ 1.
Note that the incompatibility of this scheme with
conventional gauge transformations
means that care is required with standard manipulations.
For example,
adding a boundary term 
$\int dx d\La/dx \equiv \int dx \prt_\mu \La dX^\mu/dx$
to the action \rf{standard}
would correspond to a conventional gauge transformation
and would therefore change the physics in this scheme.

\begin{figure}
\centerline{\psfig{figure=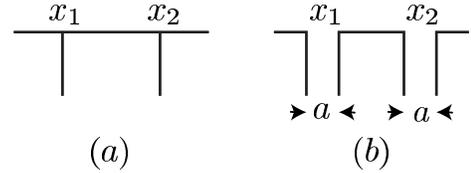,width=0.7\hsize}}
\smallskip
\caption{Regularization schemes. 
(a) Particle couplings: $\De x$ arbitrary.
(b) String couplings: $|\De x |>a$.}
\label{fig:Figure1}
\end{figure}

Although one might be tempted to dismiss the brick-wall regularization scheme
because it cannot be derived from a fundamental correlator defined
at all distances on the world sheet,
this feature may in fact make the scheme more relevant to string theory.
Introducing a cutoff distance $\aaa$ in the integral
is equivalent to introducing a cutoff for world-sheet momenta 
of the order of $1/\aaa$.
The uncertainty principle then implies that distances shorter than $\aaa$
cannot be distinguished.
From the perspective of external open strings coupling 
to the world-sheet boundary,
it is more natural to allow for a width of order $\aaa$ rather than zero.
This corresponds to excluding any distance smaller than $\aaa$
in calculating correlators between two strings coupling at different
external points,
and indeed coincident correlators corresponding to the same
external string have a natural separation of order $\aaa$.
From this perspective,
the first regularization scheme treats the background couplings 
as if they arose from point-particle interactions,
which may be unjustified as it 
omits a physical aspect of the background-string interactions.

The need in the first regularization scheme
for a nontrivial metric $G^{jk}$ 
to ensure the existence of the action and the corresponding lagrangian
${\cl}_\infty$
is a disadvantage.
The metric $G^{jk}$ must be invertible
to avoid introducing physics beyond that contained in the beta functionals.
Although this invertibility may hold for sufficiently small field values,
$G^{jk}$ can become singular under certain circumstances.
For example,
noninvertibility could occur in Eqs.\ \rf{Gval} 
for field values of order $1/g$,
suggesting that ${\cl}_\infty$ 
is inappropriate to describe the corresponding physics.
Nontrivial vacua are associated with field values of order $1/g$,
so investigating the corresponding physics with ${\cl}_\infty$ 
would appear to require considerable caution. 

An interesting argument in favor of the brick-wall scheme
arises by comparing the two results obtained above
with string field theory
\cite{ew}.
The form of the lagrangian ${\cl}_{\rm SFT}$ for this theory
comparable to ${\cl}_\infty$ and $\cl$
can be extracted from Ref.\ \cite{ks}.
We find
\bea
{\cl}_{\rm SFT} &=& 
\half\ppp(\prt^2+1)\ppp
- \frac 14 F_{\mu\nu}F^{\mu\nu}
+ \frac 1 {3!} g\check \ppp^3 
+ \half g \check \ppp \check A_\mu \check A^\mu
\nonumber\\
&&
+g \check \ppp 
( \prt_\nu \check A^\mu \prt_\mu \check A^\nu 
+ \check A^\mu \prt_\mu \prt_\nu \check A^\nu ) 
\quad ,
\label{S3}
\eea
where as before $g$ is the on-shell tree-level three-tachyon coupling.
In this expression,
the fields $f_n$ with mass $n$ 
are smeared over the distance $\sqrt{\al'}$
in the interaction terms:
$f_n \to \check f_n\equiv \exp[\al'\ln(3\sqrt3/4)
(\prt_\mu\prt^\mu - n)]f$.
The full string field theory  
is gauge invariant \cite{ew},
and the leading-order gauge transformations 
of ${\cl}_{\rm SFT}$ exhibit modifications 
\cite{ks}
corresponding to those in Eq.\ \rf{gaugeinv2}.
By construction,
${\cl}_\infty$, $\cl$, ${\cl}_{\rm SFT}$
all coincide on-shell in the transverse gauge
and hence produce the same on-shell amplitudes.

It is striking that the lagrangians $\cl$ and ${\cl}_{\rm SFT}$
agree term-by-term, 
excluding the smearing factors in ${\cl}_{\rm SFT}$.
This suggests that within the sigma model 
the use of the brick-wall regularization scheme 
may be appropriate.
We anticipate that the missing smearing factors
could also emerge in the sigma model approach 
if the positions of the attached strings
(e.g., $x_1$ and $x_2$ in Fig.\ 1(b))
are taken as having an intrinsic uncertainty of width $a$.
However, 
this issue lies beyond our present scope.

A natural question is whether appropriate redefinitions
of $\ppp(k)$ and $A^\mu(k)$
might convert lagrangians of the form
$\cl$ and ${\cl}_{\rm SFT}$
into ones of the form ${\cl}_\infty$, 
or perhaps might be used to eliminate entirely 
the static $\ppp^3$ and $\ppp A_\mu A^\mu$ terms 
and similar ones appearing at higher order.
In $\cl$,
for example,
this can indeed be done for small field values at order $g$. 
Although it is conceivable that for small enough field values 
a similar redefinition could be implemented locally
at higher order in $g$,
the procedure fails globally
whenever nontrivial extrema of the static potential exist
because the necessary field redefinitions become singular.
If indeed the presence of cubic (or higher-order) interactions 
creates extrema of the static potential away from zero, 
then the inappropriate use of field redefinitions in this context
would amount to preordaining the irrelevance of nontrivial string vacua.

\smallskip

This work is supported in part by 
the US DOE,
by NATO grant CRG-960693,
and by the Portuguese FCT.

\end{multicols}

\begin{references}
\bibitem{fn1}
An analogous regularization scheme has previously been used 
for the closed-string case
\cite{yw}.
\bibitem{fn2}
Nontrivial metric coefficients have also been encountered in
Ref.\ \cite{es}.
\end{references}
\end{document}